\documentclass{article}

\usepackage{arxiv}
\usepackage[utf8]{inputenc} 
\usepackage[T1]{fontenc}    
\usepackage{booktabs}       
\usepackage{amsfonts}       
\usepackage{nicefrac}       
\usepackage{microtype}      
\usepackage{lipsum}		
\usepackage{amsmath,amssymb}
\usepackage{graphicx}
\usepackage{color}

\newcommand{\grl}{    {Geophys. Res. Lett.}}
\newcommand{\jgr}{    {J. Geophys. Res.}}
\newcommand{\ssr}{    {Space Sci. Rev.}}

\newcommand{\apj}{ {Astrophys. J. }}
\newcommand{\apjl}{    {Astrophys. J. Lett.}}

\newcommand{\prl}{    {Phys. Rev. Lett.}}

\newcommand{\blue}{\textcolor{black}}

\def\XXint#1#2#3{{\setbox0=\hbox{$#1{#2#3}{\int}$}
     \vcenter{\hbox{$#2#3$}}\kern-.5\wd0}}

\title{Double layers in the Earth's bow shock}

\author{
  {Jiepu Sun}\\
  Space Sciences Laboratory\\
  University of California at Berkeley\\
  CA 94720, USA\\
  \texttt{11912118@mail.sustech.edu.cn} \\
   
   \And
 {Ivan Y. Vasko} \\
 Space Sciences Laboratory\\
  University of California at Berkeley\\
  CA 94720, USA\\
  \texttt{vaskoiy@berkeley.edu} \\
    
    \And
    {Stuart D. Bale}\\
      Space Sciences Laboratory\\
  University of California at Berkeley\\
  CA 94720, USA\\
    \And
    {Rachel Wang}\\
      Space Sciences Laboratory\\
  University of California at Berkeley\\
  CA 94720, USA\\

    \And
    {Forrest S. Mozer}\\
      Space Sciences Laboratory\\
  University of California at Berkeley\\
  CA 94720, USA\\
}

\begin{document}
\maketitle

\begin{abstract}
We present Magnetospheric Multiscale observations of electrostatic double layers in quasi-perpendicular Earth's bow shock. These double layers have predominantly parallel electric field with amplitudes up to 100 mV/m, spatial widths of 50--700 m, and plasma frame speeds within 100 km/s. The potential drop across a single double layer is 2–7\% of the cross-shock potential in the de Hoffmann-Teller frame and occurs over the spatial scale of ten Debye lengths or one tenth of electron inertial length. Some double layers can have spatial width of 70 Debye lengths and \blue{potential drop up to} 30\% of the cross-shock potential. The \blue{electron temperature variation} observed across double layers is roughly consistent with their potential drop. While electron heating in the Earth's bow shock \blue{occurs predominantly due to} the quasi-static electric field in the de Hoffmann-Teller frame, these observations show that \blue{electron temperature can also increase} across Debye-scale electrostatic structures.
\end{abstract}

{\bf Key points:}
\begin{enumerate}
    \item The first identification of ion-acoustic double layers and \blue{electron temperature variation} across them in the Earth's bow shock.\\ 
    
    \item Double layers have typical spatial width around ten Debye lengths or one tenth of electron inertial length.\\
    
    \item Typical potential drop across a double layer is 2--7\% of the cross-shock potential in the de Hoffmann-Teller frame.
\end{enumerate}


\section{Introduction}

Electron heating remains one of the not entirely resolved problems in the physics of collisionless shock waves \cite{Scudder95,Krasnoselskikh:ssr13,Gedalin20:apj}. Previous spacecraft measurements showed that the electron heating across quasi-perpendicular Earth's bow shock is mainly determined by the \blue{quasi-static} cross-shock electrostatic field in the de Hoffmann-Teller frame \cite{Scudder95,Hull01,Lefebvre07}. This electrostatic field is rarely possible to be measured directly, but based on the observed electron heating we expect it to have typical amplitude of a few mV/m and spatial scale intermediate \blue{between electron and ion inertial lengths} \cite{Bale05:ssr,Schwartz11,Krasnoselskikh:ssr13,Wilson2021:front}. The major effect of high-frequency \blue{fluctuations produced by plasma instabilities} in the Earth's bow shock is assumed to be a smoothing of the electron distribution function shaped by \blue{quasi-static} magnetic and electric fields \cite{Scudder95,Hull01,Lefebvre07}. There are currently experimental \cite{Chen18:prl,Wilson2021:front,Vasko22:grl} and theoretical \cite{Gedalin20:apj,Tran20,Kamaletdinov22} indications that high-frequency \blue{fluctuations} not only smooth the electron distribution function, but can also contribute to the electron heating. In this study we present experimental evidence that a fraction of \blue{the electron heating} in the Earth's bow shock \blue{can} occur across Debye-scale electrostatic double layers. \blue{Note that the term {\it heating} used in this letter is  equivalent of {\it temperature increase}, because we observe spatial profiles of the electron temperature, rather than its temporal evolution.}

It is noteworthy that double layers in the form of more or less unipolar spikes in the parallel (magnetic field-aligned) electric field are universally observed during the energy release processes in space plasma \cite{Andersson&Ergun12}. Double layers were originally reported in the auroral region \cite{Temerin82,Mozer98:grl}, while later studies showed they had plasma frame speeds around local ion-acoustic speed and spatial widths of about ten Debye lengths \cite{Ergun01:prl,Andersson02:phpl}. Similar ion-acoustic double layers were observed around plasma injection fronts in the inner magnetosphere \cite{Deng10:jgr,Malaspina14:grl}, fast plasma flows in the plasma sheet \cite{Ergun09:prl,Yuan22:grl}, and reconnection current sheets in the magnetosphere \cite{Wang16:grl,Oieroset21:phpl}. Parker Solar Probe measurements have recently revealed double layers, presumably of ion-acoustic type, in the Venusian bow shock \cite{Malaspina20:grl}. \blue{The spacecraft measurements showed that ion-acoustic double layers can indeed accelerate or heat electrons \cite{Ergun09:prl,Yuan22:grl}. The presence} of electrostatic double layers in the Earth's bow shock was pointed out in the past \cite{Hobara08,Goodrich18:iaw}, but no in-depth analysis of their properties and the associated electron heating was carried out.

In this letter we present Magnetospheric Multiscale measurements of electrostatic double layers in the Earth's bow shock and demonstrate that they are of ion-acoustic type. We show that \blue{electron temperature variation} across the double layers is roughly consistent with the potential drop across them. The origin of these double layers is discussed.

\section{Observations\label{sec2}}

We first consider a quasi-perpendicular Earth's bow shock crossing by Magnetospheric Multiscale (MMS) spacecraft around 07:56:30 UT on 4th November 2015 that was previously used to study electrostatic solitary and ion-acoustic \blue{waves \cite{Wang21:jgr,Vasko22:grl}. We use} burst mode measurements of the DC-coupled magnetic field at 128 S/s (samples per second) resolution provided by Digital and Analogue Fluxgate Magnetometers \cite{Russell16}, electron and ion moments and velocity distribution functions provided at 0.03 and 0.15 s cadence respectively by the Fast Plasma Investigation instrument \cite{Pollock16}, electric field fluctuations along with voltage signals of individual voltage-sensitive probes at 8,192 S/s resolution provided by Axial Double Probe \cite{Ergun16} and Spin-Plane Double Probe \cite{Lindqvist16}, \blue{and magnetic field fluctuations at 8,192 S/s resolution provided by the Search Coil Magnetometer \cite{LeContel16}}. The electric field was measured by two pairs of probes in the spin plane, with opposing probes mounted on tips of 60 m antennas, and a pair of axial probes mounted on tips of 14.6 m antennas along the spin axis. The electric field components in the orthogonal coordinate system associated with the \blue{probes, whose schematic can be found in Supporting Materials (SM), were} computed as $E_{12}=(V_{2}-V_{1})/120\;{\rm m}$, $E_{34}=(V_{4}-V_{3})/120\; {\rm m}$, $E_{56}=(V_{6}-V_{5})/29.2\;{\rm m}$, and then multiplied by frequency response factors to be specified below. The electric and magnetic fields will also be presented in the Geocentric Solar Ecliptic (GSE) coordinate system whose $z-$axis essentially coincides with the spacecraft spin axis.

\begin{figure}
\includegraphics[width=0.8\linewidth]{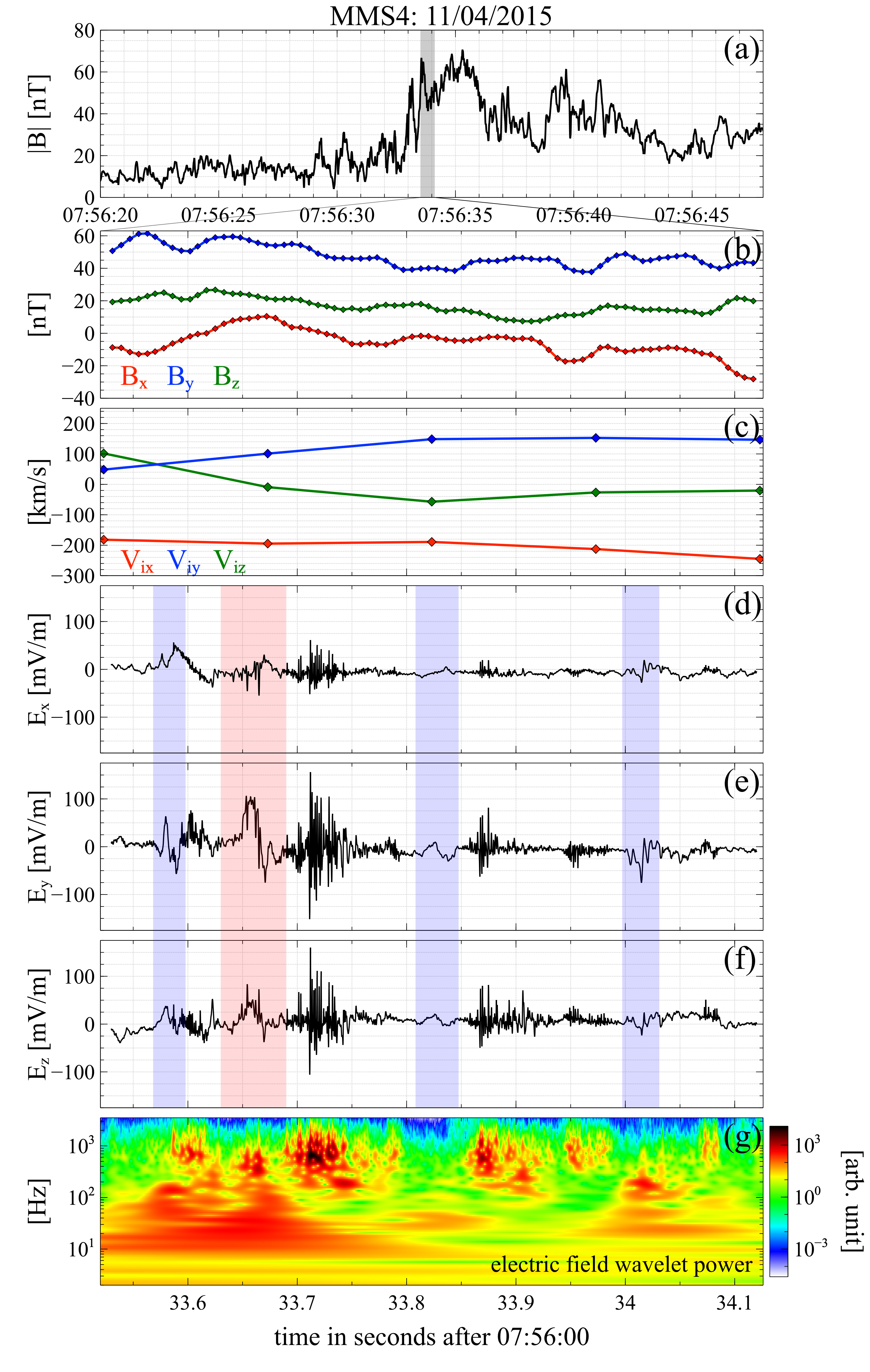}
\caption{MMS4 observation of double layers in a quasi-perpendicular Earth's bow shock crossing on 4th November, 2015. Panel (a) presents the magnetic field magnitude observed across the shock along with a highlighted region, whose expanded view is shown in the following panels: (b), (c) magnetic field and plasma flow velocity in the GSE coordinate system; (d)--(f) three electric field components measured at 8,192 S/s in the GSE coordinate system and (g) their omnidirectional continuous Morse wavelet spectrum in arbitrary units. There are sporadic bursts of high-frequency electrostatic fluctuations (above about 200 Hz) preceded by lower-frequency electrostatic structures (below 200 Hz). These electrostatic structures are highlighted in panels (d)--(f) with the largest-amplitude one highlighted in red.}
\label{fig1}
\end{figure}

Figure \ref{fig1} presents MMS4 measurements in the aforementioned Earth's bow shock crossing. Panel (a) presents the magnetic field magnitude and highlights a region around the ramp, where double layers were observed. The expanded view of that region is shown in following panels. Panel (b) demonstrates that the magnetic field is predominantly along the $y-$axis and, thus, almost within the spacecraft spin plane. Panel (c) shows that the plasma flows at about 100 km/s parallel to the local magnetic field and 200 km/s perpendicular to it. Three electric field components in panels (d)--(f) and their omnidirectional wavelet spectrum in panel (g) show a sporadic occurrence of electric field fluctuations at frequencies above a few hundred Hz. These are electrostatic ion-acoustic waves considered in the previous study \cite{Vasko22:grl}. The intriguing, though not universal, feature is that each burst of ion-acoustic waves is preceded by a solitary structure, whose typical frequency is below 200 Hz. These solitary structures are electrostatic, since no magnetic counterparts were observed, with the electric field predominantly along the $y-$axis which is parallel to the local magnetic field. We only present in-depth analysis of the solitary structure with the largest electric field amplitude, because similar analysis of other structures revealed they were of identical nature.

\begin{figure}
\includegraphics[width=0.8\linewidth]{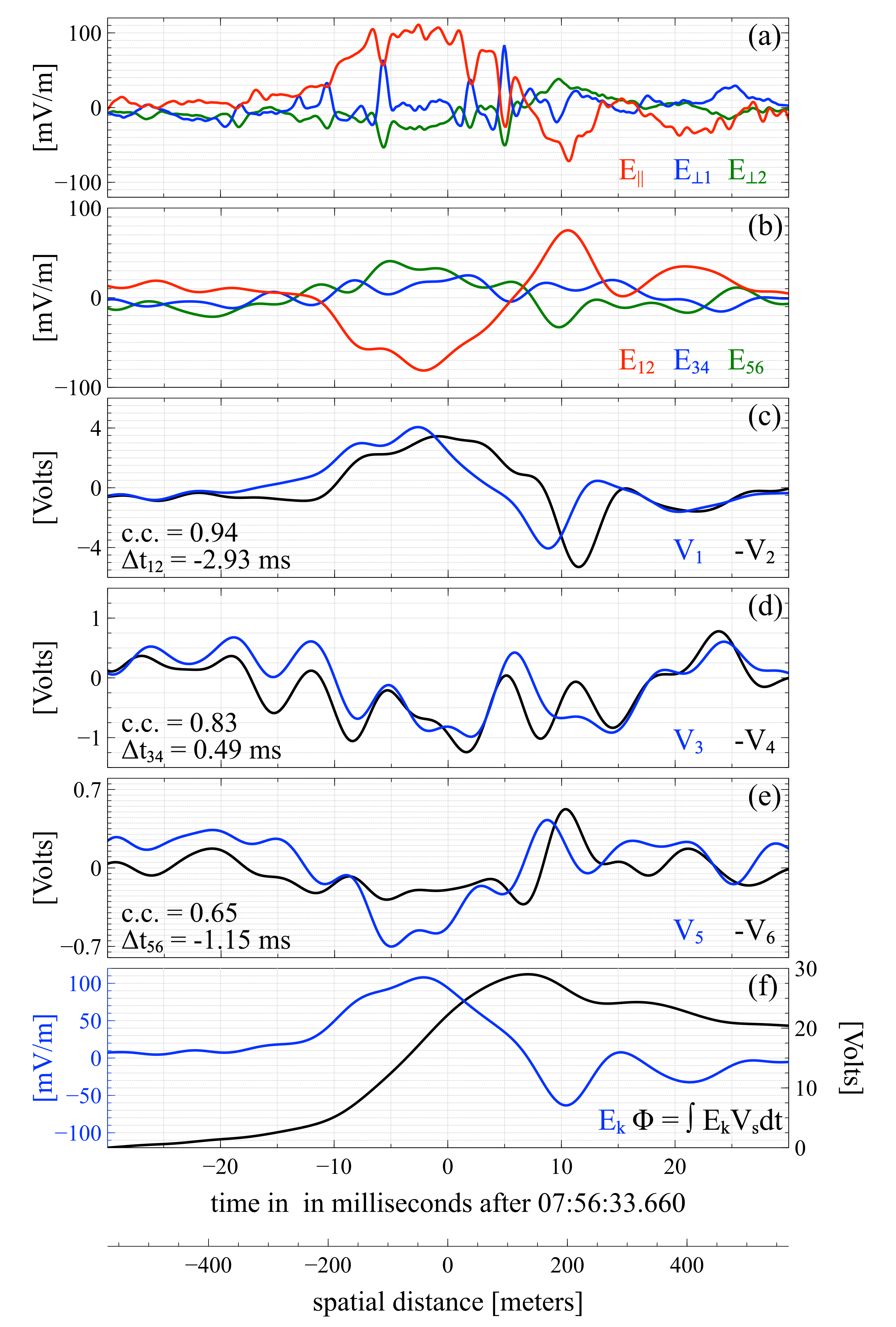}
\caption{The interferometry analysis of the double layer highlighted in red in Figure \ref{fig1}: (a) the electric field measured at 8,192 S/s in the magnetic field-aligned coordinate system (components parallel and perpendicular to the local magnetic field are shown); (b) the electric field in the coordinate system associated with voltage-sensitive probes and computed using voltage signals $V_{1}$--$V_{6}$ of probes in panels (c)--(e): $E_{12}=1.35\cdot (V_{2}-V_{1})/120\;{\rm m}$, $E_{34}=1.35\cdot(V_{4}-V_{3})/120\;{\rm m}$, and $E_{56}=1.2\cdot (V_{6}-V_{5})/29.2\;{\rm m}$; (c)--(e) low-pass filtered (cutoff frequency of 200 Hz) and offset eliminated voltage signals of four probes in the spacecraft spin plane ($V_1$--$V_{4}$) and of two probes mounted on the axial antenna along the spin axis ($V_{5}$ and $V_{6}$); cross-correlation coefficients and time delays between voltage signals of opposing probes are indicated in panels; (f) the electric field $E_{k}$ along the polarization direction \blue{$\hat{\bf k}$} computed by applying the Maximum Variance Analysis to the electric field in panel (b) and the electrostatic potential computed as $\Phi=\int E_k V_s dt$, where $V_{s} \approx 19$ km/s is the double layer speed in the spacecraft frame estimated using the time delay between $V_{1}$ and $-V_{2}$. The estimated speed $V_{s}$ allowed translating the temporal axis into the spatial axis shown at the bottom.}
\label{fig2}
\end{figure}

Figure \ref{fig2} presents an interferometry analysis of the largest-amplitude solitary structure highlighted in Figure \ref{fig1}. The electric field in the magnetic field-aligned coordinate in panel (a) demonstrates that the solitary structure has a full temporal width of 20 ms and predominantly parallel electric field with amplitude of 100 mV/m. \blue{Note electrostatic fluctuations with the temporal scale of about 1 ms, whose electric field is rather oblique to the electric field of the solitary structure.} These high-frequency fluctuations can be excluded by a low-pass filter with the cutoff frequency of 200 Hz. Panel (b) presents electric field components $E_{12}$, $E_{34}$, and $E_{56}$ computed using low-pass filtered voltage signals and frequency response factors of 1.35 and 1.2 for spin plane and axial antennas respectively. The axial antenna's factor may actually vary between 0.8 and 1.6 \cite{Wang21:jgr}, but our results are not sensitive to its specific value, because the electric field of the solitary structure is essentially in the spin plane; moreover, it is directed from probe 2 to probe 1. Applying the Maximum Variance Analysis \cite{Sonnerup&Scheible98} to the electric field in panel (b), we determined a unit vector \blue{$\hat{\bf k}$} along the electric field polarization direction; \blue{in the coordinate system associated with the probes we have $\hat{\bf k}\approx (-0.91,0.12, 0.39)$ that is within 20$^{\circ}$ of the local magnetic field}. Low-pass filtered voltage signals are presented in panels (c)--(e) along with correlation coefficients and time delays between signals of opposing probes. The highest correlation and the largest time delay were observed between $V_{1}$ and $-V_{2}$. The speed of a locally planar electrostatic structure can be estimated as follows \cite{Vasko20:front,Wang21:jgr}: $V_{s}=\hat{k}_{ij}l_{ij}/\Delta t_{ij}$, where $\Delta t_{ij}$ is the time delay between signals $V_{i}$ and $-V_{j}$ of a pair of opposing probes, while $l_{ij}$ is the corresponding antenna length. Taking into account that \blue{$\hat{k}_{12} \approx -0.91$}, $\Delta t_{12} \approx -2.93$ ms, and $l_{12}=60$ m, we obtained the speed of the solitary structure in the spacecraft frame, $V_{s} \approx 19$ km/s. Note that the electric field polarization direction \blue{$\hat{\bf k}$}, intrinsically ambiguous by 180$^{\circ}$, was chosen to be consistent with propagation from probe 1 to probe 2 as observed in panel (c). Voltage signals of other pairs of opposing probes are not perfectly correlated and corresponding time delays \blue{are less reliable}. The estimated speed allowed translating temporal profiles into spatial ones and computing the electrostatic potential of the solitary structure, $\Phi=\int E_{k}V_{s}dt$, where $E_{k}$ is the electric field along the polarization direction. Panel (f) shows that the solitary structure has a net potential drop across itself of $\Delta \Phi \approx 25$ V; thus, it is actually a double layer.

The spatial width of the double layer defined as $w = \Delta \Phi/{\rm max}(E_{k})$ is about 240 m, which is 15$\lambda_{D}$ or 0.2$\lambda_{e}$ in units of local Debye length or electron inertial length. With the spatial width defined that way, the electrostatic potential of the double layer can be approximated as $\Phi(X)\approx 0.5\Delta \Phi \left(1+{\rm tanh}(2X/w)\right)$, where $X$ is the spatial coordinate along the electric field polarization direction. \blue{Note that the spatial extent of the double layer in the plane perpendicular to the polarization direction is less than 10 km, since the double layer was not observed aboard the other MMS spacecraft spatially separated from MMS4 by at least 10 km in that plane.} We estimated the speed of the double layer in the plasma frame as $V_{s}^{*}=|V_{s}-\hat{\bf k}\cdot {\bf V}_{i}|$, where ${\bf V}_{i}$ \blue{is the plasma flow velocity}. We found $V_{s}^{*}\approx 90$ km/s or about 0.9$c_{\rm IA}$, where $c_{\rm IA}=(T_{e\parallel}/m_{p})^{1/2}\approx 100$ km/s is a rough estimation of the local ion-acoustic speed ($m_{p}$ is the proton mass). Thus, this double layer is of ion-acoustic type. The potential drop of 25 V is about 24\% of the local electron parallel temperature and about 6\% of the entire cross-shock potential $\Delta \Phi_{\rm HT}$ corresponding to the electric field in the de Hoffmann-Teller frame; the cross-shock potential of $\Delta \Phi_{\rm HT} \approx 400$ V was estimated using the electron \blue{momentum balance} (SM).

\begin{figure}
\includegraphics[width=0.8\linewidth]{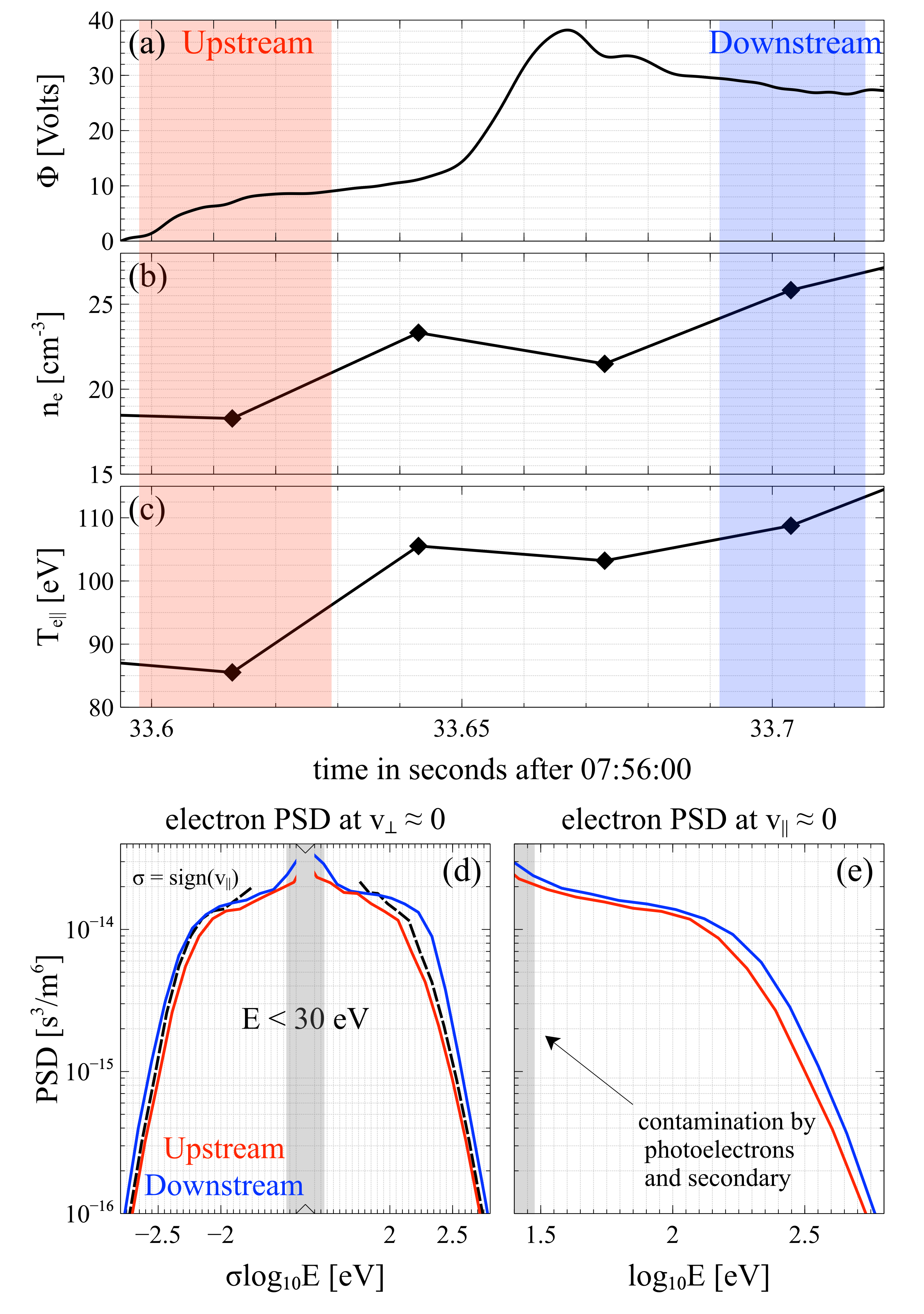}
\caption{The analysis of the electron temperature variation across the double layer shown in Figure \ref{fig2}: (a) the electrostatic potential $\Phi=\int E_k V_s dt$ obtained by integrating the low-pass filtered (with cutoff frequency of 200 Hz) electric field $E_{k}$ over an interval about twice as long as the interval shown in Figure \ref{fig2}; (b) and (c) the electron density and the parallel temperature measured \blue{at 30 ms cadence}; (d) and (e) the phase space density (PSD) of electrons collected upstream and downstream of the double layer over 30 ms intervals highlighted in panels (a)--(c). PSDs correspond to electrons with pitch angles within $(0^{\circ},30^{\circ})$ and $(150^{\circ},180^{\circ})$ in panel (d) and within $(60^{\circ},120^{\circ})$ in panel (e). \blue{The dashed line in panel (d) corresponds to the expected PSD downstream of the double layer that was computed using Liouville mapping of the upstream PSD (the energy of upstream electrons was increased by 25 eV, while their phase space density was conserved). The fluxes of electrons below about 30 eV were contaminated by secondary and photoelectrons.}}
\label{fig3}
\end{figure}

Figure \ref{fig3} presents an analysis of \blue{the electron temperature variation and momentum balance} across the double layer. Panel (a) shows the electrostatic potential obtained by integrating the low-pass filtered electric field over about 120 ms time interval, which is twice longer than that in Figure \ref{fig2}. Panels (b) and (c) present electron density $n_{e}$ and parallel temperature $T_{e||}$ over the same interval. Because the double layer can only be considered isolated over this interval (Figure \ref{fig1}e), there is merely one point of plasma measurements available upstream and downstream of it. The 30 ms time intervals where upstream and downstream electron distribution functions were collected are highlighted in panels (a)--(c). \blue{The projection of the Ohm's law onto the local magnetic field indicates that the electrostatic field of the double layer} must be balanced by the electron pressure gradient (e.g., \cite{Kelley72})
\begin{eqnarray}
e\nabla_{\blue{||}} \Phi\approx \nabla_{\blue{||}} (n_{e}T_{e\parallel})/n_{e},
\label{eq:Ohms}
\end{eqnarray}
\blue{where we omitted negligible inertia terms and neglected anomalous resistivity; the latter is a reasonable assumption in the Earth's bow shock \cite{Schwartz88,Schwartz21:jgr,Lefebvre07}}. The electron density increases across the double layer from 18 to 25 cm$^{-3}$, while parallel electron temperature increases from 87 to 107 eV. Therefore, the expected potential drop is $\Delta \Phi\approx \Delta (n_{e}T_{e\parallel})/e\langle n_{e}\rangle\approx 50$ V, where $\langle n_{e}\rangle\approx 21.5$ cm$^{-3}$ is the averaged electron density. The Ohm's law prediction is consistent with the observed potential drop of 25 V within a factor of two. The analysis of electron distribution functions upstream and downstream of the double layer reveals the reason for the quantitative disagreement. 

\blue{Panel (d) presents} the distribution function of upstream and downstream electrons streaming quasi-parallel to the local magnetic field line. In accordance with the presence of a net potential drop, the downstream distribution function is wider than the  upstream one. \blue{It is wider however than the expected distribution computed via Liouville mapping of the upstream distribution. The expected distribution in panel (d) was obtained by increasing the energy of upstream electrons by 25 eV, while keeping their phase space density conserved. Note that the distribution of downstream electrons could not be computed below about 50 eV, because upstream electron fluxes below about 30 eV were contaminated by secondary and photoelectrons (e.g., Gershman et al. \cite{Gershman17:jgr}). In addition, the distribution of downstream electrons below 25 eV would not be possible to compute using the Liouville mapping, because these electrons are trapped downstream of the double layer; see a schematics of the electron phase space in the SM.} We do not expect any widening of the distribution function of electrons streaming quasi-perpendicular to the local magnetic field, but observations in panel (e) demonstrate the opposite. \blue{This is a strong indication} of temporal variations that occurred within 90 ms between observations upstream and downstream of the double layer. The temporal variations are not necessarily local, because each distribution function was collected over 30 ms, which \blue{translates} into spatial resolution of a few hundred kilometers ($>50$ eV electrons cover at least 100 km over 30 ms). We believe that the quantitative inconsistency within a factor of two between the Ohm’s law prediction and the observed potential drop is caused by temporal variations, which resulted in parallel temperature \blue{increase} and \blue{widening of the distribution function} larger than expected for the local potential drop of 25 V.

\begin{table}[h]
\caption{The table presents various properties of double layers shown in Figures \ref{fig2} and \ref{fig4}: spacecraft frame speed $V_{s}$ and plasma frame speed $V_{s}^{*}$ along with its value in units of local ion-acoustic speed estimation, $c_{\rm IA}=(T_{e\parallel}/m_{p})^{1/2}$; spatial width $w$ in physical units as well as in units of local Debye length $\lambda_{D}$ and electron inertial length $\lambda_{e}$; potential drop $\Delta \Phi$ in physical units as well as in units of local parallel electron temperature $T_{e\parallel}$ and in units of cross-shock potential $\Delta \Phi_{\rm HT}$ in the de Hoffmann-Teller frame. \label{table}}
    \begin{tabular}{c|c|c|c|c|c|c|c|c}
     &$V_s$ \& $V_{s}^{*}$ [km/s]&$w$ [m]&$\Delta\Phi$ [V]&$w/\lambda_D$ &$w/\lambda_e$&$e\Delta\Phi/T_{e\parallel}$&$V_s^*/c_{\rm IA}$&$\Delta\Phi/\Delta\Phi_{\rm HT}$\\
     \hline
     Figure \ref{fig2}&19 \& 90& 240&25&15&0.23&24\%&0.9&6\%\\
     \hline
     Figure \ref{fig4}a&52 \& 104& 90&7.5&8&0.1&8\%&1.1&2\%\\
     \hline
     Figure \ref{fig4}b&51 \& 62& 670&32&67&0.9&33\%&0.6&32\%\\
     \hline
     Figure \ref{fig4}c&27 \& 50& 55&3&9&0.07&9\%&0.9&5\%\\
     \hline
     Figure \ref{fig4}d&12 \& 95& 65&2.2&13&0.1&7\%&1.8&7\%
    \end{tabular}
\end{table}

\begin{figure}
\includegraphics[width=1.0\linewidth]{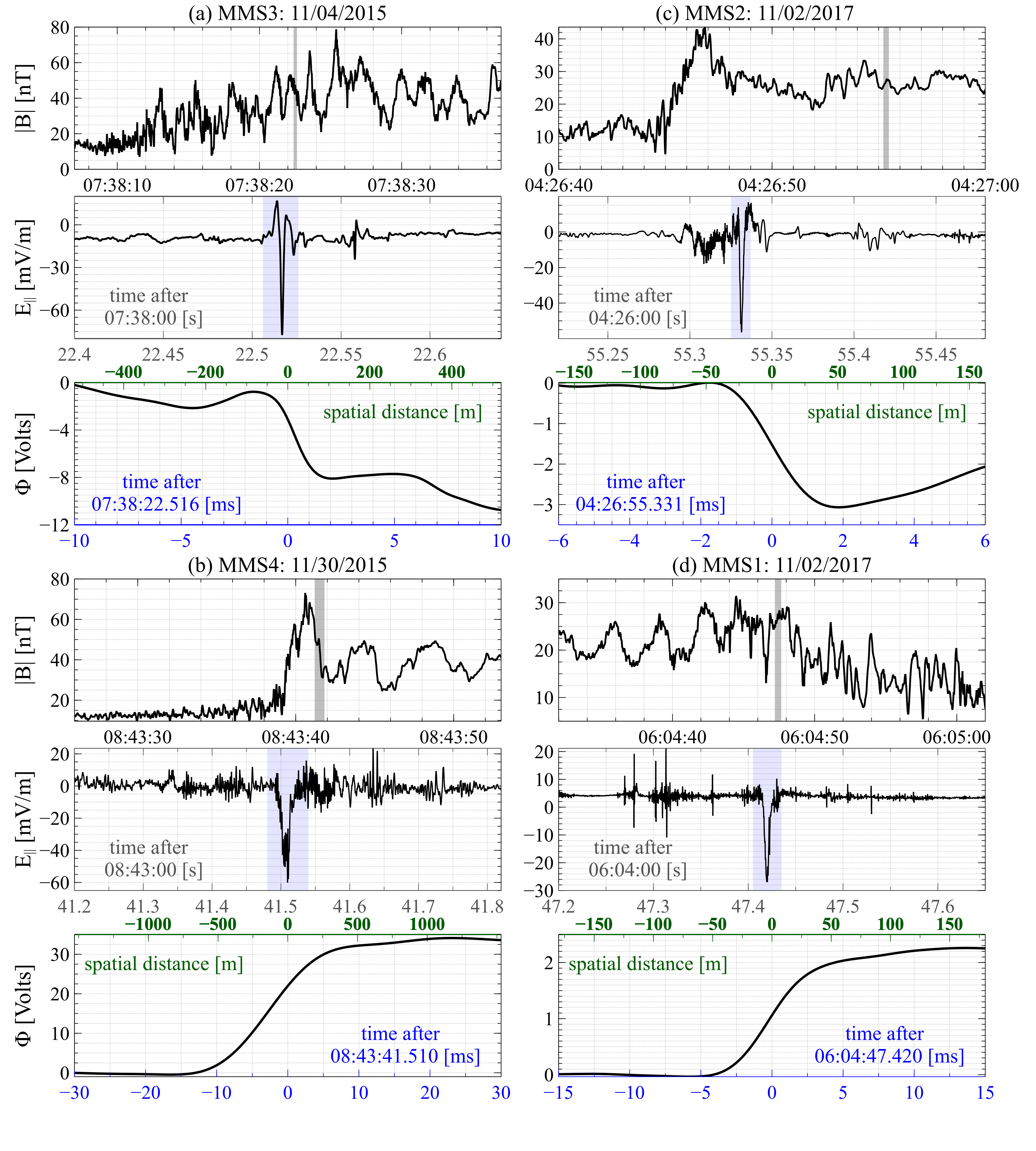}
\caption{Double layers observed in other quasi-perpendicular Earth's bow shock crossings. In each set of panels (a)--(d), the upper panel presents the magnetic field magnitude observed across a shock, the middle panel shows the parallel electric field over the interval highlighted in the upper \blue{panel}, and the bottom panel presents the electrostatic potential computed using the double layer speed estimated using interferometry (SM). Electrostatic potentials are shown over time intervals highlighted in the middle panels; the spatial axis corresponding to the temporal axis is at the top of the bottom panels. \blue{Table \ref{table} summarizes various parameters of the double layers}.}.
\label{fig4}
\end{figure}

Figure \ref{fig4} demonstrates double layers observed in several other Earth's bow shock crossings considered previously by Wang et al. \cite{Wang21:jgr}. The upper panel in each panel set (a)--(d) presents the magnetic field magnitude \blue{along} with a highlighted region where the double layer was observed, while the middle panel presents the parallel electric field over that interval. These panels demonstrate that double layers can occur within or close to the shock ramp as well as in the downstream region. The electric fields of all double layers are parallel to the local magnetic field within 20$^{\circ}$ (SM). We carried out interferometry analysis to estimate speeds and electrostatic potentials of the double layers (SM). Bottom panels in (a)--(d) demonstrate electrostatic potentials, while various parameters of the double layers are presented in Table \ref{table}. These double layers have plasma and spacecraft frame speeds within 100 km/s, spatial widths within 50--700 m, and potential drops up to 30 V. They are of ion-acoustic type, because their plasma frame speeds are comparable with local ion-acoustic \blue{speed}, $V_{s}^{*}/c_{\rm IA}\approx 0.5$--$2$. The typical spatial width of the double layers is 8--15$\lambda_{D}$ or 0.07--0.23$\lambda_{e}$, though some double layers can have spatial width up to 70$\lambda_{D}$ or 0.9$\lambda_{e}$. The potential drop across a single double layer is typically 2--7\%, though can be up to 30\%, of the cross-shock potential in the de Hoffmann-Teller frame. In units of local parallel electron temperature, these potential drops vary between 7\% and 30\%. For all the double layers the prediction of the Ohm's law (\ref{eq:Ohms}) is consistent within 30\% with the observed potential drop; the only exception is the double layer in panel (d), whose potential drop is relatively small and the Ohm's law estimation is most likely dominated by temporal variations (SM).

\section{Discussion and Conclusion}

Previous spacecraft measurements showed that electron heating in the Earth's bow shock was determined by the quasi-static electric field in the de Hoffmann-Teller frame \cite{Scudder95,Hull01,Lefebvre07}, which arises from different electron and ion dynamics \cite{Goodrcih&Scudder84}. The typical spatial scale of the quasi-static electric field is between electron and ion inertial length \cite{Bale05:ssr,Schwartz11,Krasnoselskikh:ssr13,Wilson2021:front}. While we believe that electron heating in the Earth's bow shock is indeed dominated by the quasi-static electric field, in this letter we demonstrated that \blue{the electron temperature can also increase} across electrostatic double layers. The \blue{electron temperature} variation across a single double layer can be 2--7\% of the cross-shock potential in the de Hoffmann-Teller frame and occurs over spatial scale \blue{of only} about ten Debye lengths or one tenth of electron inertial length. Some double layers can have spatial width of 70 Debye lengths and \blue{potential drops} up to 30\% of the cross-shock potential. \blue{Importantly, double layers can have their high-potential side facing magnetosheath (Figures \ref{fig2} and \ref{fig4}b) as well as solar wind (Figures \ref{fig4}a,c,d). In the former the electron temperature across double layers increases toward magnetosheath, while in the latter it increases toward solar wind. Note that the quasi-static electric field always results in electron temperature increase toward the magnetosheath.}

Double layers presented in this letter are electrostatic and they fundamentally differ from electric field spikes reported in the foot of a quasi-perpendicular Earth's bow shock \cite{Chen18:prl}; those spikes \blue{were} nonlinear whistler mode structures \cite{Balikhin97,Walker99:grl}. Similar electrostatic double layers were previously reported in the auroral region \cite{Ergun01:prl,Andersson02:phpl}, plasma sheet \cite{Ergun09:prl,Yuan22:grl}, inner magnetosphere  \cite{Deng10:jgr,Malaspina14:grl}, reconnection current sheet \cite{Wang16:grl,Oieroset21:phpl}, and Venusian bow shock \cite{Malaspina20:grl}. In rare cases the double layer propagation velocity was revealed by interferometry \cite{Ergun01:prl,Andersson02:phpl,Yuan22:grl}, while typically it was assumed to coincide with the local ion-acoustic speed. Our interferometry analysis showed that double layers in the Earth's bow shock have plasma frame speed in the range of 0.5--2$c_{\rm IA}$, where we used a rough estimation of the local ion-acoustic speed, $c_{\rm IA}=(T_{e\parallel}/m_{p})^{1/2}$, because its exact value depends on non-Maxwellian features of electron and ion distribution functions (e.g., Vasko et al. \cite{Vasko17:grl}). These double layers are of ion-acoustic type and, thus, identical with double layers in the auroral region \cite{Ergun01:prl,Andersson02:phpl} and plasma sheet \cite{Yuan22:grl}. \blue{Similarly to previous observations, the double layers are associated with high-frequency electric field fluctuations (Figures \ref{fig1} and \ref{fig4}), though that is not a universal feature (Figure \ref{fig4}a). The origin of these high-frequency fluctuations is not apparent, because the double layers are rather weak to produce electron beams at the high-potential side (Figure \ref{fig3}).}

\blue{The causal relation between the double layers and the temperature variation across them is not obvious. Double layers formed by a current-driven ion-acoustic instability can indeed result in electron heating \cite{Sato&Okuda81:jgr,Newman01:prl,Vazsonyi20}, but double layers formed at a contact boundary between cold and hot electron populations result in {\it no actual electron heating} \cite{Li&Drake12:apj,Li&Drake14:apj}. In the case of a current-driven ion-acoustic instability in a uniform plasma \cite{Sato&Okuda81:jgr}, a double layer is associated with a negative potential spike (ion hole). Although ion holes are abundant in the Earth's bow shock \cite{Vasko18:grl,Wang20:apjl,Wang21:jgr}, the observed double layers are not associated with them (Figures \ref{fig2} and \ref{fig4}) and, hence, fundamentally different from ones observed in simulations by Sato and Okuda \cite{Sato&Okuda81:jgr}. The double layers are similar to those produced by a current-driven instability developing around a localized plasma density perturbation \cite{Newman01:prl}, but they are also not different from double layers formed at a contact boundary between cold and hot electron populations, the latter produced by some other heating mechanism \cite{Li&Drake12:apj,Li&Drake14:apj}. The origin of the double layers need to be established in the future to reveal their ability in actual electron heating.}

In conclusion, we showed that \blue{the electron temperature} in the Earth's bow shock \blue{can increase} across Debye-scale electrostatic structures. The estimation of net contribution of these structures to the electron heating remains a challenge and the mechanism producing these double layers remains to be revealed.

\section{Open Research}
All the additional information on double layers considered in this letter can be found in the Supporting Materials. We thank the MMS team for excellent data available at \url{https://lasp.colorado.edu/mms/sdc/public/about/browse-wrapper/}.

The work of J.S. and R.W. was supported by National Science Foundation grants No. 2026680. The work of I.V. and F.M. was supported by NASA Heliophysics Guest Investigator grant No. 80NSSC21K0730. I.V. also thanks the International Space Science Institute, Bern, Switzerland for supporting the working group "Resolving the Microphysics of Collisionless Shock Waves". I.V. thanks Anton Artemyev and Sergey Kamaletdinov for valuable contributions.

\bibliographystyle{unsrt}  


\end{document}